\documentclass{article}
\usepackage[final]{neurips_2023}

\usepackage{microtype}
\usepackage{graphicx}
\usepackage{subfigure}
\usepackage{booktabs} 
\usepackage{listings}
\usepackage{algorithm}
\usepackage{algorithmic}
\usepackage{hyperref}

\usepackage{amsmath}
\usepackage{amssymb}
\usepackage{mathtools}
\usepackage{amsthm}

\usepackage[capitalize,noabbrev]{cleveref}

\theoremstyle{plain}

\theoremstyle{definition}

\theoremstyle{remark}

\usepackage[textsize=tiny]{todonotes}

\title{\textit{nbi}: the Astronomer's Package for Neural Posterior Estimation}

\author{%
Keming Zhang$^{1,2}$\thanks{Correspondence to: kemingz@berkeley.edu.} \thanks{Eric and Wendy Schmidt AI in Science Postdoctoral Fellow} \quad Joshua S. Bloom$^{2}$ \quad St\'efan van der Walt$^{2}$ \quad Nina Hernitschek$^3$\\
$^1$University of California, San Diego \quad $^2$University of California, Berkeley\\ \quad $^3$Universidad de Antofagasta\\}

\begin{document}

\maketitle

\begin{abstract}
Despite the promise of Neural Posterior Estimation (NPE) methods in astronomy, the adaptation of NPE into the routine inference workflow has been slow. We identify three critical issues: 
the need for custom featurizer networks tailored to the observed data, the inference inexactness, and the under-specification of physical forward models. To address the first two issues, we introduce a new framework and open-source software \textit{nbi} (\textit{Neural Bayesian Inference}), which supports both amortized and sequential NPE. First, \textit{nbi} provides built-in ``featurizer'' networks with demonstrated efficacy on sequential data, such as light curve and spectra, thus obviating the need for this customization on the user end. Second, we introduce a modified algorithm SNPE-IS, which facilities asymptotically exact inference by using the surrogate posterior under NPE only as a proposal distribution for importance sampling. These features allow \textit{nbi} to be applied off-the-shelf to astronomical inference problems involving light curves and spectra. We discuss how \textit{nbi} may serve as an effective alternative to existing methods such as Nested Sampling. Our package is at \url{https://github.com/kmzzhang/nbi}.

\end{abstract}

\section{Problem Statement}

With observations $x$, model parameters $\theta$, and a prior belief on $p(\theta)$, the forward model informs the data generating process (i.e.\ the likelihood) $p(x|\theta)$, and the inverse problem is to solve for the posterior distribution of model parameters $p(\theta|x)$ given the observations\footnote{Both $x$ and $\theta$ are vectors}. In this paper, we are concerned with inference problems with tractable likelihood functions, where the parameter space has finite dimension and each parameter is bounded. This type of problem is common in astronomy, and contrasts with problems with a nuisance parameter space that is difficult or impossible to marginalize over, which is common is cosmology and particle physics and have been the original focus of simulation-based inference techniques (e.g., \citealt{weyant_likelihood-free_2013,cameron_approximate_2012,hahn_approximate_2017,hsu_improving_2018}).

Because none but the simplest forward models are invertible, inference is usually approached using methods such as MCMC and Nested Sampling, at the core of which are iterative comparisons between model predictions and observations.
In the analysis of large survey data, the simulations produced during the modeling and inference of one target usually cannot be applied to the analysis of subsequent targets.
For computationally expensive forward models, the reuse of simulations across inference instances is thus desired, a computational approached referred to as amortization.

Central to the technique of Neural Posterior Estimation (NPE) is the training of a conditional Neural Density Estimator (NDE) to obtain an inverse model---a neural surrogate posterior $q(\theta|x)$.
The NDE is usually trained using the negative-log-likelihood (NLL) loss on pairs of ($x, \theta$) generated from the forward model, which leads to the minimization of the Kullback–Leibler (KL) divergence from the NDE surrogate posterior to the exact posterior (e.g., \citealt{zhang_real-time_2021}). Downstream inference then becomes a forward pass through the trained NDE, which incurs minimal additional cost. This is referred to as Amortized NPE (ANPE), which allows for the inference of large number of datasets at constant/sub-linear costs.

While ANPE benefits the batch analysis of large numbers of observed targets, Sequential NPE (SNPE) aims to accelerate the inference of one given target by using active learning. 
Therefore, the use cases of SNPE is similar to those ordinarily solved with existing methods such as MCMC.
Compared to ANPE, SNPE starts off with a much smaller training set sampled from the prior for the first round. After proper training, the NDE surrogate posterior for the target of interest is then used to generate parameters for the next round training set. SNPE then proceeds over multiple rounds until the surrogate posterior converges to the exact posterior (for that particular target). By producing more simulations around the final posterior, SNPE is more simulation efficient.
Nevertheless, the inference of a new target would require retraining from the first round, or the second round if they share the same prior. In practice, therefore, there would be a tipping point where sufficient number of inference tasks would warrant switching from SNPE to ANPE.

The efficacy of NPE for astronomical inverse problems has been demonstrated for a wide range of problems over the years, including galaxy SED fitting \citep{hahn_accelerated_2022,khullar_digs_2022}, gravitational waves \citep{dax_real-time_2021}, binary-lens microlensing \citep{zhang_real-time_2021}, and exoplanet atmosphere retrieval \citep{vasist_neural_2023}. Despite the growing popularity and maturity of NPE, it has yet to become part of the astronomer's routine data-analysis toolbox such as MCMC. We enumerate three critical issues that have slowed the integration of NPE into our domain research workflow.

The first impediment is the need for custom featurizer/embedding networks tailored to the observed data, which compress the high-dimensional observed data into a low-dimensional vector to be ``read in'' by the NDE as the conditional.
This need for customization beyond standard convolutional and recurrent neural networks creates a steep learning curve for the domain expert less steeped in deep learning.
On the other hand, there exist network architectures that are well benchmarked and demonstrated to be effective for common astronomical datasets (e.g.\ \citealt{becker_scalable_2020,zhang_classification_2021,chan_convolutional_2021,moreno-cartagena_positional_2023}, notably including sequential data like light curves and spectra.

The second aspect is the scientist's common skepticism of results produced by the ``neural-network black box.'' In response, NPE application papers often conduct accuracy and calibration tests to demonstrate the trained model as being trustworthy. While these exercises may be useful for ANPE, they are less straightforward for SNPE. An adequate (and preferably intuitive) metric that validate the SNPE-produced results is thus required for SNPE to become a practical and competitive alternative to existing algorithms such as Nested Sampling.

The third aspect is the out-of-distribution (OOD) problem. Our forward models are often simplistic representations of the actual data-generating process provided by nature and our data-collecting instruments. The fact that the training and testing data comes from different distributions means that any good behavior as demonstrated from the aforementioned accuracy and calibration tests \textit{may} break
\footnote{To clarify, the OOD problem is caused by and not equivalent to the model under/mis-specification problem. Under mis-specified models, both the maximum a posteriori (MAP) solution and the full Bayesian posterior are still defined. The OOD problem, which is specific to machine-learning methods, may prevent NPE from even producing this wrong (but sometimes useful) answer.}
down when applied to the real dataset. Fortunately, the OOD problem only affects a subset of inverse problems. The current work aims to alleviate the first two aspects, leaving the third aspect for future work (e.g., \citealt{wang_sbi_2023}).

\section{The \textit{nbi} Framework}

To address the first two issues, we introduce a new NPE framework and software called \textit{Neural Bayesian Inference (nbi)}, which is designed as a generic inference tool with out-of-the-box functionality for certain astronomical problems.
Our package follows the scikit-learn API.\\\\
\begin{lstlisting}
engine = nbi.NBI(flow, featurizer, simulator, priors)
engine.fit(n_sims=1000, n_rounds=1, n_epochs=100, noise=noise)
y_pred, weights = engine.predict(x_obs, x_err, n_samples=2000)
\end{lstlisting}
Our framework is similar to existing frameworks such as \textit{sbi} \citep{tejero-cantero_sbi_2020} and \textit{LAMPE}\footnote{https://github.com/francois-rozet/lampe} in many ways. We support both ANPE and SNPE (for $n_{\rm rounds}>1$), and we utilize the Masked Autoregressive Flow (MAF; \citealt{papamakarios_masked_2017}) as the NDE, whose hyper-parameters are specified via the dictionary \texttt{flow} during initialization. The MAF is adapted from our previous work \citep{zhang_real-time_2021}, which truncates the scaling factor for each MAF block to improve training stability. On the other hand, our framework differentiates from existing frameworks with the following features:

\textbf{Pre-Built Featurizers}: We provide featurizer network architectures that are known to work well for certain types of astronomical data. For the initial release, the ResNet-GRU architecture adapted from our earlier work \citep{zhang_real-time_2021} is supplied to allow for out-of-the-box use with light curves and spectra data.
The featurizer network is initialized simply by specifying its hyper-parameters in the \texttt{featurizer} keyword as a dictionary. Our package documentation provides guidance on the selection of these few parameters.
A wider range of pre-built featurizer networks will be included in future releases to support more data types.

\textbf{Splitting up the Simulator}:
Forward models in astronomy can often be separated into computationally cheap and expensive parts. At the minimum, the cheap part contains the measurement noise but may also include flux normalization, PSF convolution, etc.
The expensive part is supplied through \texttt{simulator}, which is called during \texttt{engine.fit()} to produce the training set, which is saved to disk. The cheap part is supplied through \texttt{noise}, which is applied as data augmentation during training.

\textbf{Asymptotically Exact Inference}: The availability of explicit and tractable likelihood functions for a wide range of inference problems in astronomy allows one to re-weight the NPE output---the surrogate posterior---into an asymptotically exact posterior via importance sampling. We will discuss the integration of importance sampling into \textit{nbi} in the next section.

\section{Asymptotically Exact Inference}

The availability of tractable likelihood functions for vast astronomical inference problems contrasts sharply with problems traditionally tackled with simulation-based inference. At the core of \textit{nbi} is the use of the likelihood function to re-weight the NDE surrogate posterior into asymptotically exact posteriors via importance sampling \citep{paige_inference_2016,muller_neural_2019,dax_neural_2023}. 
The importance weight for each discrete sample of the NDE surrogate posterior is:
\begin{equation}
    w_i\propto \dfrac{p(\theta_i|x_{\rm obs})}{q(\theta_i|x_{\rm obs})}.
\end{equation}
where $\theta_i$ is the $i$-th NDE posterior sample, $p(\theta_i|x_{\rm obs})=p(x_{\rm obs}|\theta_i)\cdot p(\theta_i)$ is the exact posterior probability calculated via the likelihood and prior, and $q(\theta_i|x_{\rm obs})$ is the probability of $\theta_i$ under the NDE surrogate posterior. The importance weights are normalized to a sum of one and are returned along with the posterior samples with \texttt{NBI.predict()}.

To evaluate the quality of the surrogate posterior, we avoid obscure measures such as the KL-divergence, and opt for the effective sample size that allows for a straightforward interpretation:
\begin{equation}
\label{eq:eff}
    n_{\rm eff} = \dfrac{1}{\sum_i w_i^2} - 1.
\end{equation}

The effective sample size is an estimate of the sample size required to achieve the same level of precision if that sample was drawn from the exact posterior, as opposed to the NDE surrogate posterior.
Note that we have defined Equation \ref{eq:eff} such that its minimum value is zero, instead of the ordinary one.
If the NDE posterior is exact, then the effective sample size would be near the total sample size.
If the NDE posterior is way off, then the effective sample size would be zero, indicating that the inference results should not be accepted.
The ratio between the effective and total sample size represents the simulation efficiency\footnote{For context, the simulation efficiency for Metropolis-Hasting MCMC is typically below 10\%, considering an optimal acceptance rate of 23.4\% and correlation in the parameter chains.}.

Because the NDE posterior is only used as a proposal distribution to importance sampling, the user can trust the \textit{nbi} output to the extent indicated by the effective sample size. In this way, the ``estimation'' aspect of Neural Posterior Estimation is removed, and its asymptotically-exact nature therefore allows NPE to serve similar use cases regularly solved with MCMC and Nested Sampling. After the training phase, the user could then utilize the NDE posterior solely for importance sampling to reach their desired final effective sample size.

\textbf{SNPE}: The integration of importance sampling with NPE is particularly interesting for SNPE. Under SNPE, the NDE surrogate posterior is trained over multiple rounds, where the surrogate posterior from the previous round is used to generate the training set of the next round. Here, the compute expended for simulating the next round training set can be directly used for calculating the importance weights for the current round NDE posterior. Thus, each round of SNPE contributes to the final effective sample size at no extra cost. The ratio between the effective and total sample size (sampling efficiency) for each round becomes an indicator for the convergence of the NDE posterior towards the exact posterior, and thus an early stopping criterion. Once the surrogate posterior stops improving for the next round, one may directly acquire more effective posterior samples simply by importance sampling from the previous round surrogate posterior with the highest sampling efficiency. The full SNPE with importance sampling integration (SNPE-IS) algorithm is presented in Algorithm 1.

\begin{algorithm}[tb]
\caption{SNPE with Importance Sampling (SNPE-IS)}
\label{alg:nbi}
\begin{algorithmic}[1]
\STATE {\bfseries Input:} observed data $x$, simulator $f$, likelihood $p(x|\theta)$, prior $p(\theta)$, number of rounds $R$, simulations per round $N$
\STATE {\bfseries Output:} posterior samples $\{\theta_{1:N\cdot R}\}$, weights $\{w_{1:N\cdot R}\}$, effective sample sizes $\{n_{\rm eff,1:R}\}$, final surrogate posterior $\hat{q}(\theta)$
\STATE
\STATE Set proposal to prior $\hat{q}(\theta)\leftarrow p(\theta)$ and $n_{\rm eff, 0}=0$
\FOR{$i=1$ {\bfseries to} $R$}

\STATE draw $\{\theta_{1:N}\}\sim \hat{q}(\theta)$
\STATE simulate $\{x_{1:N}\} \leftarrow f(\theta_{1:N})$
\STATE calculate weights $\{w_{1:N}\}$ with Equation 1
\STATE calculate $n_{\rm eff,i}$ with Equation 2
\IF{$n_{\rm neff, i} < n_{\rm eff, i-1}$}
\STATE revert $\hat{q}(\theta)$ to previous round
\STATE {\bfseries Exit:} NDE stopped improving
\ENDIF
\STATE Train $\hat{q}(\theta)$ on $\{\theta_{1:N},x_{1:N}\}$ using Negative-Log-Likelihood (NLL) loss
\ENDFOR
\end{algorithmic}
\end{algorithm}

The integration of importance sampling also facilitates the use of the simple NLL loss across all rounds. A property of the NLL loss is that the NDE would converge to a posterior with the training parameter distribution as the prior. Thus, the NLL loss is typically restricted to ANPE and the first round of SNPE where the training set is indeed produced from the prior.
For the second round of SNPE onward, the Automatic Posterior Transformation (APT) \citep{greenberg_automatic_2019} loss is commonly applied.
On the high-level, the APT loss is a re-weighted version of the NLL loss such that the NDE posterior converges as if the training set is drawn from the prior, regardless of the distribution that the training parameters are drawn from.
For any given simulated observation within a training batch, APT calculates the probability under the NDE for not just the parameter vector corresponding to that simulation, but on all parameter vectors in the entire training batch, which APT uses to re-weight loss.
While the APT loss allows the training set drawn from any proposal distribution to be re-weighted back to the specified prior, it does so at the computational expense of $O(B^2)$, where $B$ is the batch size. Thus, the APT loss becomes expensive for large-batch training.

Since asymptotic exactness is already facilitated by the use of importance sampling, for \textit{nbi} we opt to use the simple NLL loss for all SNPE rounds. This would be perhaps most problematic near convergence, where $q(\theta|x_{\rm obs})\sim p(\theta|x_{\rm obs})$. The next round NDE would therefore converge to $\sim p(\theta|x_{\rm obs})^2$, with overestimated precision. However, this degradation is reflected in the reduced effective sample size and sampling efficiency for the subsequent round, which therefore can be used as a criterion to terminate training. At this point, one may use the previous round surrogate posterior for importance sampling to achieve a final desired effective sample size.

\section{Discussion}

We consider the current version (v0.4.1) of \textit{nbi} ``production-ready'' for inference problems involving sequential data, such as light curves and spectra.
The only customization we require is a ``noise'' function that captures the instrumental effects of the underlying dataset.
In a companion paper \citep{zhang_stellar_2023}, we apply \textit{nbi} to the problem of stellar spectra fitting, and discuss how \textit{nbi} facilitates easy the distribution of the trained amortized inference models.
Looking forward, we plan to include more types of featurizer networks to allow for the out-of-the-box utility of \textit{nbi} to be extended to more types of data. One aspect worth mentioning is specialized networks for irregularly sampled light curve. We point out that for single inference tasks utilizing SNPE, there is no distinction between regularly and irregularly sampled data, and the 1D ResNet-GRU network provided with \textit{nbi} should be sufficient.
We outline several aspects for an extended version of this work.

\textbf{NLL vs.\ APT loss for SNPE}: We argued for the use of basic NLL loss in SNPE due to the favorable computational scaling to the batch size. In future work, we will also implement the APT loss in \textit{nbi}. We expect the APT loss to lead to better surrogate posteriors and thus larger effective sample sizes for every round, but at increased training computational cost. We will provide practical recommendations of deciding between the two.

\textbf{Horizontal Benchmarking} against existing NPE frameworks such as \textit{sbi} and LAMPE. Even though all mentioned frameworks implement similar NPE algorithms, it is likely that the performance differs. In particular, \textit{nbi} implemented a modified MAF that improves training stability. As \textit{nbi} uniquely integrates importance sampling, this benchmark will only evaluate the final NLL loss for the same task.

\textbf{Vertical Benchmarking}: The real competition for \textit{nbi}---along with other NPE codes---are the existing inference algorithms like Nested Sampling and Sequential Monte Carlo such as PocoMC \citep{karamanis2022accelerating}, all of which initialize from the Bayesian prior. Note that MCMC is not a proper baseline method because MCMC is only a sampler and technically requires first finding the \textit{maximum a posteriori} (MAP) solution, a task that is often non-trivial.

Lastly, we point out a critical issue for NPE, which is the out-of-distribution (OOD) and model mis-specification problem. Reliable uncertainty quantification is of the utmost importance in science. As such, we cannot rely on the ability of neural networks to generalize, and must seek solutions to the OOD problem with statistical guarantees.
The use of importance sampling makes it explicit when OOD is severe enough, future works should consider algorithms to improve the robustness of NPE against this issue.
The recent work of \citet{wang_sbi_2023} tackled two critical issues of out-of-distribution measurement errors and missing data. There remain many other directions for future works, such as under-specified physics (e.g., missing chemistry in spectral models).

\subsection*{Acknowledgment}
This work is supported by NSF award \#2206744: ``CDS\&E: Accelerating Astrophysical Insight at Scale with Likelihood-Free Inference.'' This project is supported by the Eric and Wendy Schmidt AI in Science Postdoctoral Fellowship, a Schmidt Futures program. KZ thanks Dan Foreman-Mackey for enlightening discussions on importance sampling.



\end{document}